\documentclass[namedreferences]{solarphysics}
\usepackage[hyperref,optionalrh,solaromanenum]{spr-sola-addons}
\usepackage{graphicx}        
\usepackage{color}           
\usepackage{comment}         
\usepackage{soul}

\newcommand{\dthta}{\Delta \theta}
\newcommand{\dlt}{\Delta t}
\newcommand{\dssn}{SSN_{27}}

\chardef\us=`\_

\begin{document}

\begin{article}
\begin{opening}

\title{Causal Analysis of Influence of the Solar Cycle and Latitudinal Solar-Wind Structure on Corotation Forecasts}

\author[addressref={aff1,aff2},corref,email={ae0221@coventry.ac.uk}]{\fnm{Nachiketa}~\lnm{Chakraborty}}
\author[addressref=aff2,email={h.turner3@pgr.reading.ac.uk}]{\fnm{Harriet}~\lnm{Turner}}
\author[addressref=aff2,email={m.j.owens@reading.ac.uk}]{\fnm{Mathew}~\lnm{Owens}}
\author[addressref=aff2,email={matthew.lang@reading.ac.uk}]{\fnm{Mathew}~\lnm{Lang}}

\address[id=aff1]{School of Computing, Electronics and Mathematics, Coventry University, United Kingdom}
\address[id=aff2]{Department of Meteorology, University of Reading, Earley Gate, PO Box 243, Reading, RG6 6BB, UK}

\runningauthor{Author-a et al.}
\runningtitle{Example paper}

\begin{abstract}
  Studying solar wind conditions is central to forecasting impact of space weather on Earth. Under the assumption that the structure of this wind is constant in time and corotates with the Sun, solar wind and thereby space weather forecasts have been made quite effectively. Such corotation forecasts are well studied with decades of observations from STEREO and near-Earth spacecrafts. Forecast accuracy depends upon the latitudinal separation (or offset $\dthta$) between source and spacecraft, forecast lead time ($\dlt$) and the solar cycle via the sunspot number (SSN). The precise dependencies factoring in uncertainties however, are a mixture of influences from each of these factors. And for high precision forecasts, it is important to understand what drives the forecast accuracy and its uncertainty. Here we present a causal inference approach based on information theoretic measures to do this. Our framework can compute not only the direct (linear and non-linear) dependencies of the forecast mean absolute error (MAE) on SSN, $\dthta$ and $\dlt$, but also how these individual variables combine to enhance or diminish the MAE. We provide an initial assessment of this with potential of aiding data assimilation in the future. 
\end{abstract}
\end{opening}

\section{Introduction}
     \label{S-Introduction} 
     
Forecasting terrestrial space weather impacts \citep[e.g. ][]{cannon_extreme_2013} necessitates knowledge of the up-stream solar wind conditions which will encounter the Earth's magnetosphere in the future. Currently, direct (in situ) solar wind observations are only routinely available near the Earth-Sun line at the first Lagrange point, L1, giving less than 40 minutes forecast lead time. Physics-based simulations of the whole Sun-Earth system can potentially provide forecast lead times of 2 to 5 days, but there remain many technical and scientific challenges to this approach \citep{luhmann_coupled_2004, toth_space_2005, merkin_predicting_2007}.   

A simple, yet robust, alternative forecast of near-Earth solar wind conditions can made using observations anywhere in the ecliptic plane by assuming the structure of the solar wind is fixed in time and corotates with the Sun. For example, observations in near-Earth space can be used to predict conditions at the same location a whole solar (synodic) rotation ahead, approximately 27.27 days \citep{bartels_twenty-seven_1934, owens_27_2013,kohutova_improving_2016}. Of course, the structure of the corona and solar wind does evolve over such time scales, particularly around solar maximum. From the L5 Lagrange point, approximately 60$^\circ$ behind Earth in its orbit, the corotation time is approximately 5 days. This is sufficiently long that the forecast lead time is useful, but sufficiently short that the corotation approximation is generally appropriate \citep{simunac_situ_2009, thomas_evaluating_2018}. Partly for these reasons, \textit{Vigil}, the upcoming  operational space-weather monitor, will make routine observations at L5 \citep{kraft_remote_2017}.

Assessing and quantifying the factors which influence the accuracy of corotation forecasts is important directly for improved corotation forecasting, but also for effective data assimilation of the solar wind observations into solar wind models \citep{lang_variational_2019, lang_improving_2021}, as it informs the expected observational errors. Longitudinal separation between the observing spacecraft and the forecast point -- and hence the forecast lead time -- is obviously expected to increase forecast error, as the steady-state assumption becomes increasingly invalid. We may also expect that this effect would be more pronounced (and corotation forecasts generally less accurate) around sunspot maximum, when the corona is known to be more dynamic and the occurrence of time-dependent coronal mass ejections (CMEs) increases \citep{yashiro_catalog_2004}. \citep[However, see ][ for evidence that this effect is reduced near the ecliptic plane]{owens_rate_2022}. Similarly, it has been argued using simulation data that corotation forecast error should increase with latitudinal separation of observing spacecraft from forecast position \citep{owens_near-earth_2019}, and that this effect is maximised at sunspot minimum \citep{owens_quantifying_2020}.

The OMNI dataset of near-Earth solar wind observations \citep{king_solar_2005} allows us to assess corotation forecasts over nearly five complete solar cycles. As near-Earth observations are used to make near-Earth forecasts one solar rotation ahead, the forecast lead time is fixed at 27.27 days and the latitudinal separation, caused by Earth's motion over a solar rotation, reaches a maximum value of around $3.5^\circ$. The twin spacecraft of the Solar-Terrestrial Relations Observatory (STEREO) \citep{kaiser_stereo_2005} provide a  means to assess the performance of corotation forecasts over a larger parameter range. The spacecraft launched into Earth-like orbits in late 2006, with STEREO-A moving ahead of Earth in its orbit, and STEREO-B behind, separating from Earth at a rate of 22.5$^\circ$ per year. This allows the corotation forecast to be assessed for a full range of longitudinal separations -- and hence forecast lead times between 0 and 27.27 days -- and, due to the inclination of the ecliptic plane to the solar equator, latitudinal separations covering the range $\pm 15^\circ$. More than a solar cycle of data is available (although the STEREO-B spacecraft was lost in 2014), allowing the effect of increasing solar activity to be estimated.

However, while uniquely valuable, assessing corotation forecasts with the STEREO dataset does present a number of challenges. Longitudinal and latitudinal separation from Earth are interdependent, as both are due to the same orbital geometry. Due to timing of launch and the orbital period, solar activity also varies approximately in step with the orbit; the spacecraft launched just before sunspot minimum and reached maximum separation just after sunspot maximum. Thus it is difficult to isolate and quantify the individual sources of error in corotation forecasting \citep{turner2021corot}. This kind of problem is ripe for causal analysis. 

Study of cause and effect is central to all branches of sciences and there are questions in solar physics -- such as factors affecting corotation forecasts -- that can be cast in those terms. In non-interventional (or observational) systems like the Sun, causal discovery is the process of inferring mechanisms or models relating cause and effects from data. But even when principal mechanisms are known from physics, causal frameworks can also be used as a diagnostic tool to determine how uncertainty in one or more variable influences another. This is very useful in making forecasts. Typically, establishing a causal relationship between variables entails determining their conditional dependency \citep{Granger1969, 2000caus.book.....P}. For random variables, both continuous and discrete, this is done via probabilistic measures. Conditional dependency has traditionally been established with Granger causality \citep{Granger1969} and these measures are mostly derived from information theory, i.e., they are `Shannon based' \citep{Schreiber2000, Kraskov2004, Williams2010}. In addition, for time-series data, the time order of events is also critical to establishing causality. Time-lags between different variables need to be carefully evaluated. Therefore the temporal resolution of time-series must be sufficient for establishing the direction of information flow; missing data can lead to spurious correlations \citep{Runge2018}. Non-linear correlations between multiple drivers can be very difficult to disentangle. We here attempt to address and demonstrate this with a framework \citep{vanleeuwen2021framework} which uses a transformed information theoretic measure that applies to both discrete and continuous variables. Typically the current state-of-the-art causal estimates are point estimates: Data is used to produce a single number to quantify the causal relationships. There is no robust uncertainty quantification. Addressing this in general, is a work in progress \citep[for eg.,][]{heckerman2020tut, Runge2018}. However, we will provide an elementary estimate of the distribution of the strength of causal relationships - the causal strength, $cs$ from hereinafter.

Our goal in this work is to provide an initial assessment of the causal dependencies between the accuracy of a forecast, the {\it target or ``effect" variable}, with the {\it driver or ``cause" variables}. For reasons explained above, the driver variables are assumed to be solar activity (quantified by sunspot number), forecast lead time (which is primarily determined by longitudinal spacecraft separation for the OMNI and STEREO observations) and latitudinal spacecraft separation. The typical approach would be to cross-correlate these variables, or rather the time-series associated with them, pairwise. However, as these relationships can often be nonlinear and multivariate we need more advanced estimators such as those based on information theoretic measures like mutual information and higher order terms \citep{ChakrabortyvanLeeuwenDMIF}. So the approach we follow here is to start with the analog of pairwise correlation, but with the non-linear estimator; mutual information. We then introduce a third variable, via conditional mutual information, to disentangle inter-dependencies amongst three driver/cause variables, in order that mediated or induced effects can be isolated. In principle, a full causal network \citep{Runge2018, vanleeuwen2021framework} can be constructed using time-series observations. But this comes with computational and, in certain situations, interpretation challenges. Hence we leave this for future work.

We describe the solar wind observations from OMNI and STEREO A and B spacecraft in section~\ref{sect:obs}. Next, we introduce the causal inference methods, demonstrating their application to the OMNI observations in section~\ref{sect:causalmethods}. We compute the distribution of causal relationships, first pairwise, quantified in terms of the mutual information using a non-linear information theoretic measure (subsection~\ref{sect:mutualinf}), examine the time averaging effect on sunspot number (subsection~\ref{sect:sunspotnumberaveragingeffect}), followed by the conditional mutual information to separate influence of the third variable (subsection~\ref{sect:cmiii}). We use 27-day corotation forecasts (also called `recurrence' or `27-day persistence' forecasts) using only OMNI data first, as it eliminates the lead-time as a variable by design ; this leaves us with testing 2 (instead of 3) drivers:the solar activity encoded in the (smoothed) Sunspot Number (or $\dssn$) and the latitudinal offset. By first learning dependencies in this simpler dataset, we then compare effects of this same subset of drivers in the STEREO datasets ignoring at first the lead time (subsection~\ref{omnistbstaconsistency}). Following this, in Section~\ref{sect:stereohigherorder}, we study induced or mediated dependencies with lead time included, by using the STEREO datasets. Finally we interpret the results and conclude whilst looking at future opportunities to improve forecasts in section~\ref{S-Conclusion}. 
 
\section{Observations}

\label{sect:obs}
   \begin{figure}    
   \centerline{\includegraphics[width=0.95\textwidth,clip=]{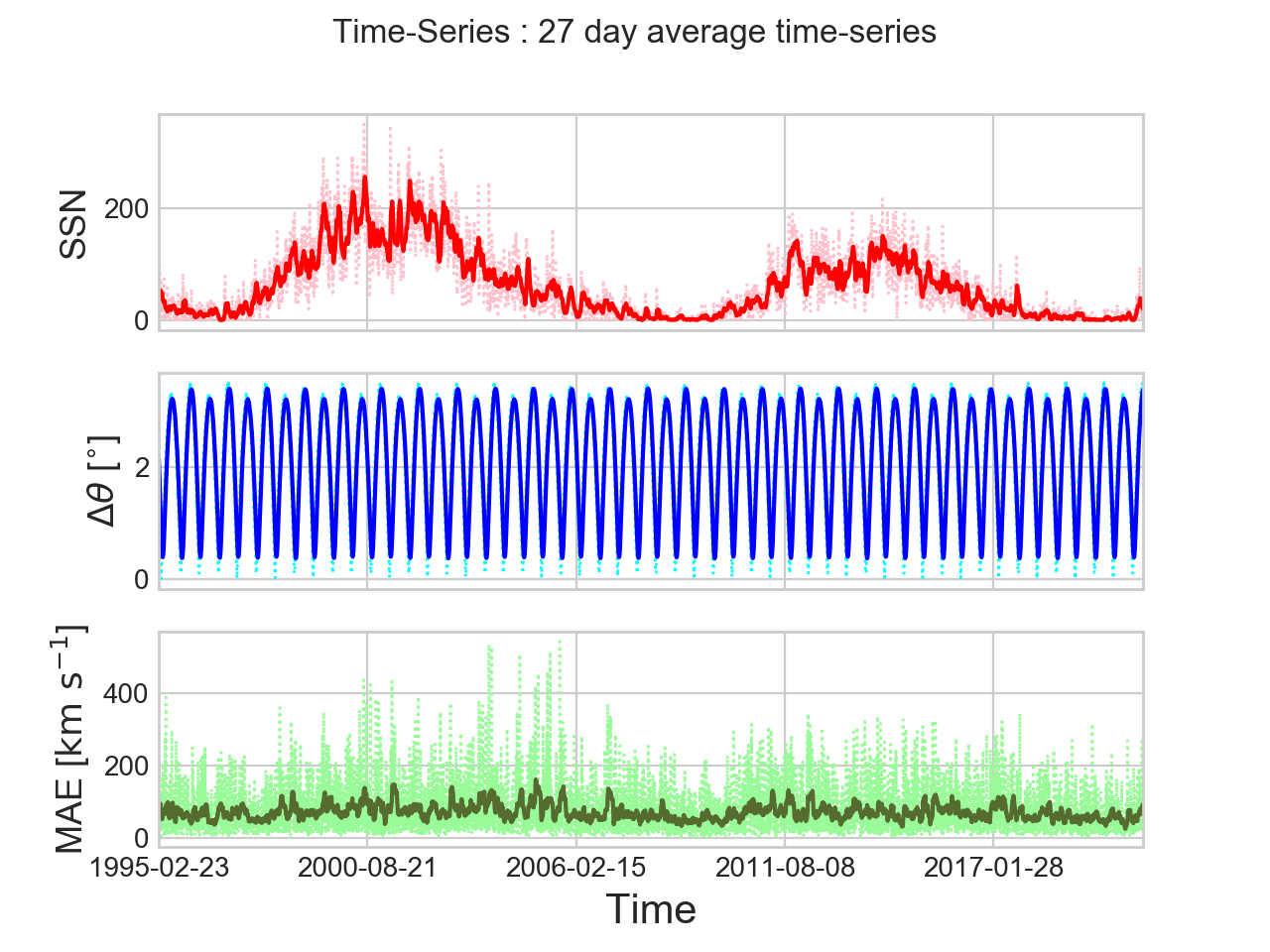}
 }
              \caption{A summary of the corotation forecast of solar wind speed obtained by using OMNI near-Earth observations to forecast near-Earth conditions 27.27 days ahead. Top: Sunspot number. Middle: The absolute value of the latitudinal separation between observation and forecast location $(\Delta \theta)$. Bottom: The mean absolute error in the solar wind speed corotation forecast. All properties are calculated at 1-day resolution (dotted lighter curves), then averaged over 27 days (solid darker curves).}
   \label{fig:fulltsOMNI}
   \end{figure}
   
      \begin{figure}    
   \centerline{\includegraphics[width=0.95\textwidth,clip=]{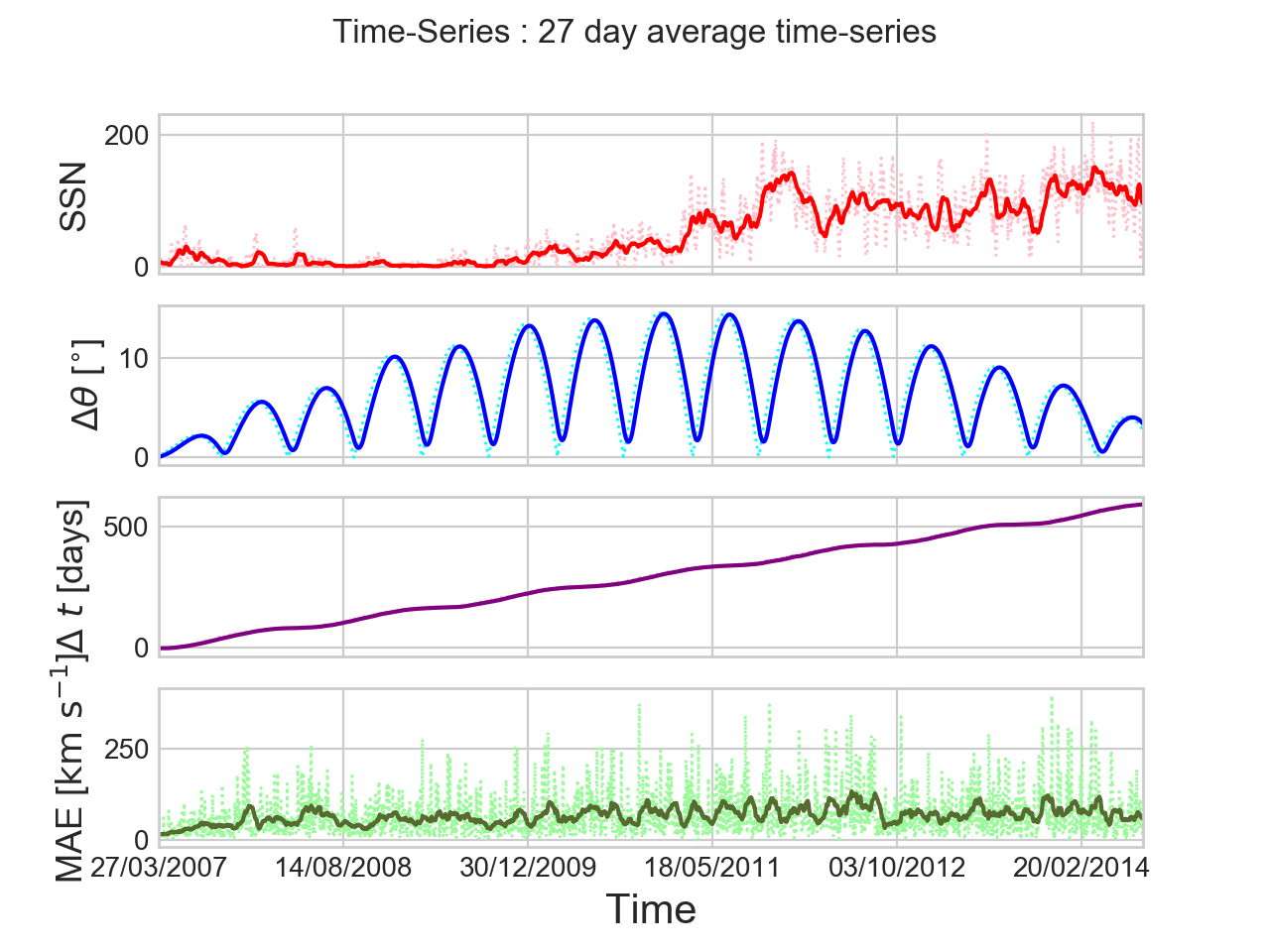}
 }
              \caption{A summary of the corotation forecast of solar wind speed obtained by using STEREO-B observations to forecast conditions at the STEREO-A spacecraft. Top: Sunspot number. Second row: The absolute value of the latitudinal separation between observation and forecast location  $(\Delta \theta)$. Third row: Forecast lead time (directly proportional to longitudinal separation, and to a much lesser extent, radial separation of spacecraft). Bottom: The mean absolute error in the solar wind speed corotation forecast. All properties are calculated at 1-day resolution (dotted lighter curves), then averaged over 27 days (solid darker curves).}
   \label{fig:fulltsSTBSTA}
   \end{figure}

Two primary data sets are used in this study. Firstly, the OMNI dataset of near-Earth solar wind conditions \citep{king_solar_2005}. Data are available from \url{https://omniweb.gsfc.nasa.gov/}. Prior to 1995, data coverage varies significantly, so the period of study is limited to 1995 to present. Secondly, the STEREO dataset, which is available from \url{https://stereo-ssc.nascom.nasa.gov/data.shtml}. STEREO-A data are used from the whole mission, 2007-present, while STEREO-B data are only available until 2014. All data are averaged to 1-day resolution to remove the effect of small-scale stochastic structure, such as waves and turbulence \citep{verscharen_multi-scale_2019}. 

Solar wind speed corotation forecasts are produced by ballistically mapping data from the observation radial distance to 1 AU, then applying a corotation delay consistent with the longitude separation. By far the dominant factor is the longitudinal separation. Further details can be found in \citet{turner2021corot}. For each forecast we compute the mean absolute error (MAE) between the forecast and observed solar wind speed.

For solar cycle context, we use the daily sunspot number (SN), provided by SILSO \citep{clette_new_2016} and available from \url{https://www.sidc.be/silso/}.

Figure \ref{fig:fulltsOMNI} shows a summary of OMNI data used to make a 27.27-day lead time forecast of near-Earth conditions. By eye, some correlation can be seen between the MAE and SN. E.g. There are few intervals of MAE above 250 km/s during the solar minima of 1996-97, 2009-10 or 2019-20. Conversely, there is no immediately obvious relation between MAE and the absolute latitudinal separation between observation and forecast location, $\Delta \theta$. However, the $\Delta \theta$ variation here is very small, arising from Earth's latitudinal orbital motion over a 27.27-day interval and reaching a maximum magnitude of around 3.5$^\circ$.

Figure \ref{fig:fulltsSTBSTA} shows the summary of STEREO-B observations used to forecast solar wind speed at STEREO-A. As the spacecraft separate in longitude, the forecast lead time, $\Delta t$, increases almost linearly. The maximum value of $\Delta \theta$ grows as the spacecraft increase their absolute longitudinal separation until mid 2010, then declines as the spacecraft move closer together (behind the Sun, from Earth's point of view). There is a somewhat linear growth in MAE from 2007 to 2012, though without further analysis it is not possible to say whether this is the result of sunspot number (post smoothing as we will see), $\Delta t$ or the amplitude of $\Delta \theta$ increasing through this time. Or some combination of those variables.
 
\section{Methods : Causal Dependencies of Corotation Forecasts}
      \label{sect:causalmethods}      
We wish to study the principle drivers of the error in the corotation forecasts. In order to do that, we perform a causal analysis on the mean absolute error (MAE) as the target/effect variable and the sunspot number ($SN$), latitudinal offset ($ |\dthta| [^\circ]$), and the forecast lead time ($\dlt$ [days]) as the principal driver/cause variables. With this setup we can use a non-linear measure of dependency to compute the causal relationships between these variables. There are a number of choices for such measures: those based on information theory as (conditional) mutual information \citep{Kraskov2004}, transfer entropy \citep{Schreiber2000}, directed information transfer \citep{amblard2009}, etc. We chose the mutual information (and its conditional variants) as it is well studied \citep[e.g.,][]{vanleeuwen2021framework, Runge2015a} and there are robust estimators available, along with an analytical result for Gaussian variables. 
The mutual information $I(x;y_{1:N})$ between a target process $x$ and a possible driver process $y$, or a whole range of driver processes denoted in our general formalism  \citep{vanleeuwen2021framework} by $y_{1:N}$ (or sometimes y, z, w,...etc.) is defined via the Shannon entropy $H(..)$ as 

   \begin{eqnarray}   
   \label{eqn:mutinf}
   I(x;y_{1:N}) &=& H(x)  - H(x|y_{1:N}) \\      
                &=& \int p(x,y_{1:N}) \log \left[\frac{p(x , y_{1:N})}{p(x)\ p(y_{1:N})}\right] \;dx dy_{1:N}
   \end{eqnarray}
Mathematically, the mutual information $I(x;y_{1:N})$ is a positive definite quantity. It can be thought of as the reduction in entropy (or uncertainty) in the target (here $x$) in presence of information content from the driver variables (here $y_{1:N}$). 
\subsection{\label{sect:vennintro}Symbolic representation : Venn Diagram Visualisation}
\begin{figure}    
   \centerline{\includegraphics[width=\textwidth,clip=]{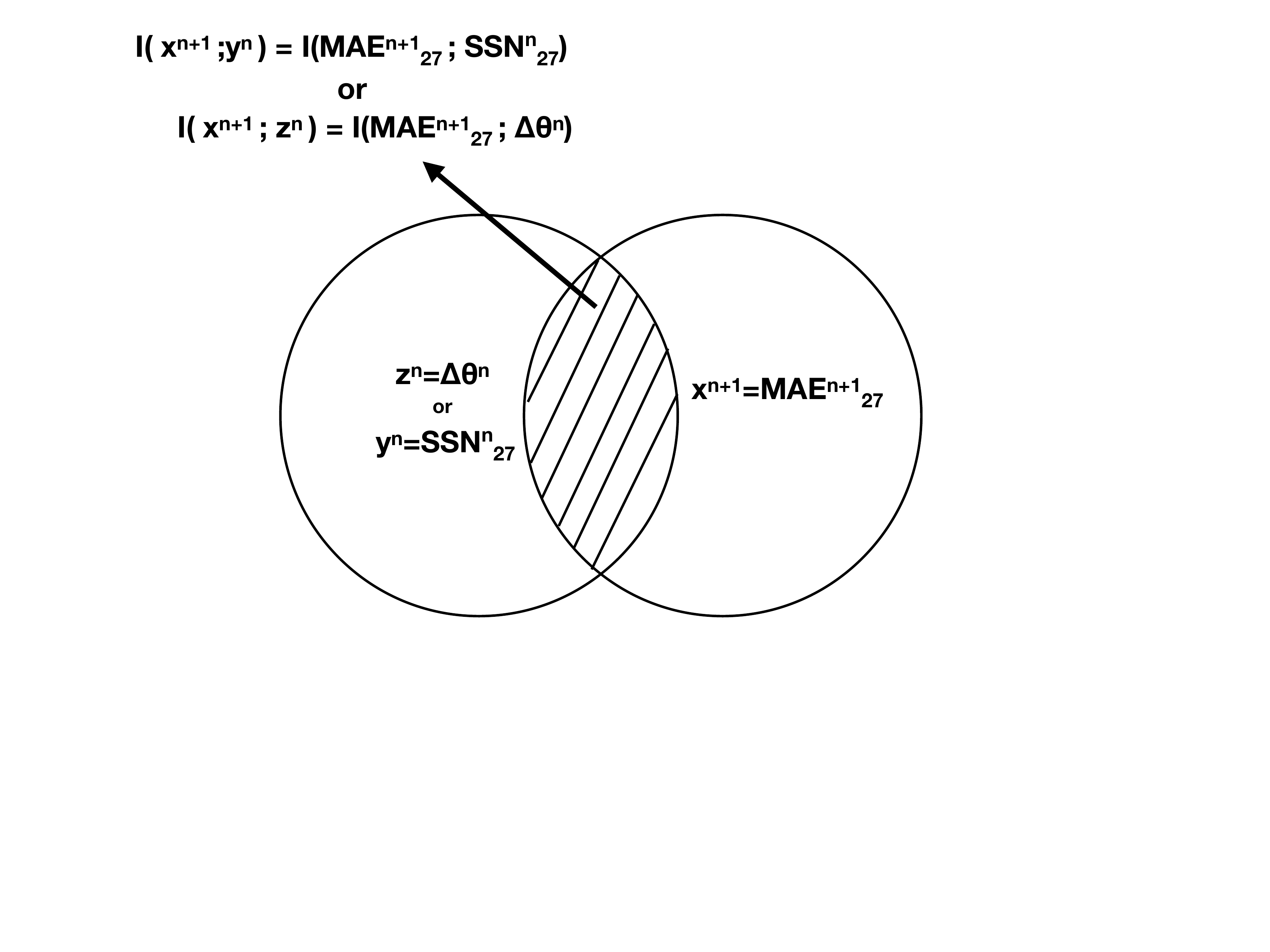}
              }
              \caption{The figure shows a graphical illustration of the mutual information via Venn diagrams of the conditional entropies $H(x|y)$ or $H(x|z)$. One circle represents the entropy content in the target - forecast accuracy (at time $n+1$) and the second represents that of one driver - either SSN or $\dthta$. The intersection represents the reduction of entropy in the target by knowledge of the driver.}
   \label{fig:mutualinfovenn}
 \end{figure}

This is shown graphically as an Information Venn Diagram in figures~\ref{fig:mutualinfovenn} (and ~\ref{fig:intvenn} for higher order terms that we will discuss later). The circles represent the conditional entropy ($H(x|y)$) of the individual variables and the intersection (shaded region with lines) represents the reduction in entropy of variable due to the presence of the other, which is the mutual information ($I(x ; y)$) defined in eqn.~\ref{eqn:mutinf}. For our application, one of these variables is the target and the other a driver ; hence the superscripts $n+1$ and $n$ showing different time indices. The labels also show the specific case on hand with the solar wind variables (MAE, SSN, $\dthta$), but we will elaborate on these in upcoming subsections.  The drivers that causally influence the target would reduce the entropy and the extent of this reduction is viewed as the extent of causal influence. On the other hand, if a driver does not have a causal influence, it does not reduce the entropy and the mutual information of the target with that driver is zero. Graphically this would mean a separation of the two circles with zero overlap. There are limitations to a formal interpretation of all situations in terms of Venn diagrams - this will become clear for higher order terms like interaction information described in subsection~\ref{sect:cmiii} \citep{GhassamiK17}. Hence, these Venn diagrams serve as a visualisation to build up our intuition rather than be a formal representation.
   
\subsection{Mutual Information : Pairwise dependency between Latitudinal Offset, Sunspot Number and MAE}
\label{sect:mutualinf}

With the goal of disentangling causal influences of drivers in corotation forecasts, we begin with OMNI data used to make a forecast at Earth. In this case, corotation forecasts have a fixed lead time of 27.27 days and forecast error, MAE, inherently has two primary drivers, the time variation of the sun -- approximated by the sunspot number (SN) -- and latitudinal offset ($\Delta \theta$) between the observation and forecast position (i.e. between Earth's location 27.27 days apart). This provides a relatively simple causal network to explore with our framework.

\begin{figure}    
     \centerline{
     \includegraphics[width=0.5\textwidth,clip=]{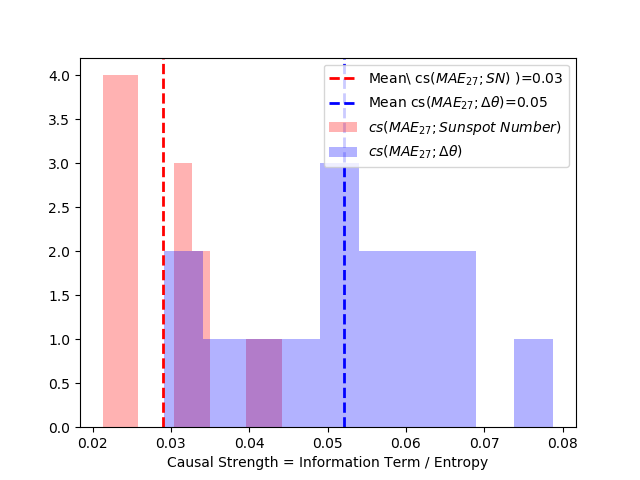}
     \includegraphics[width=0.5\textwidth,clip=]{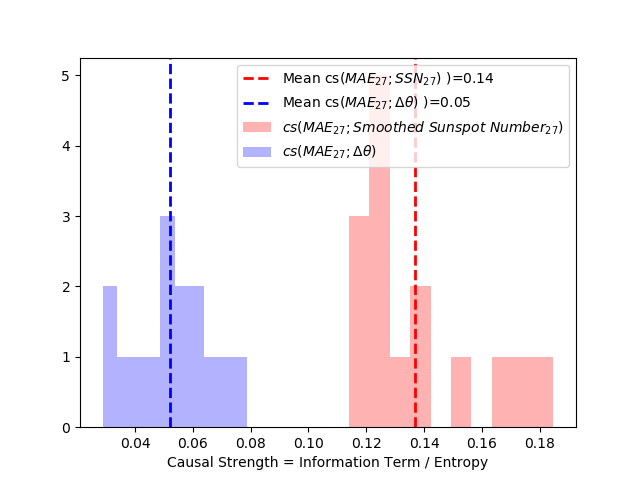}
     }
              \caption{Histograms of causal strengths, cs, of drivers -- sunspot number(SN) and latitudinal offset ($\dthta$) -- on the driver, corotation MAE obtained from 27.27-day forecasts using OMNI data. Dashed lines represent mean cs values used in causal diagrams. We take the 27-day smoothed MAE as the target in both cases. On the {\em Left:} SN and $\dthta$ are used at daily resolution ; cs computed from daily SN. {\em Right:} cs computed from 27-day smoothed SN.
              cs for SN shows a greater dependence on $|\dthta|$ than on SN, whereas the 27 day smoothed SN clearly shows the greater association of solar cycle with MAE by suppressing the stochasticity and emphasising the solar activity.}
   \label{fig:omnissnmae}
   \end{figure}

We compute the mutual information (MI) between pairs of the target and one of the drivers,  e.g., I(MAE ; $|\dthta|$). Given the length of the observation time series, we can empirically estimate the distribution of these quantities as histograms. The mutual information serves as the measure of causal dependency between pairs of one of the drivers and the target variable. Once again we refer to figure~\ref{fig:mutualinfovenn} for a visualisation graphically via Venn diagram described in section~\ref{sect:vennintro}. In this figure, the example is given for variables x and y representing target MAE, and driver either $\dthta$ or SN. In other words, we determine the reduction in entropy (or random uncertainty) in MAE, due to $\dthta$ or SN. It must be noted, that we break the symmetry between the two variables (target v/s driver), with the driver (cause) as lagging in time with respect to the target (effect). The quantities (or rather their distributions) represented by these information diagrams are estimated in figure~\ref{fig:omnissnmae}. 

As a positive definite quantity with no upper limit, MI can take very large values. Thus it is useful to normalise this measure, which is possible in a number of ways. One option is to normalise it with the total entropy or uncertainty in the variable $x$, giving the causal strength, $cs(x ; y_{1:N}) = \frac{I(x ; y_{1:N})}{H(x)}$ or simply $cs(x ; z) = \frac{I(x ; z)}{H(x)}$ for two variables : target x and driver z. There is a challenge here ; the entropy we use is for continuous variables, also known as the differential entropy, which can acquire negative values. In practice, we do not encounter this here in our applications. However, to mitigate this effect -- and for general interpretation -- we will ultimately use relative causal strengths to the total over all the drivers combined; in these relative causal strengths we ignore the contribution of noise or unmodeled drivers to merely focus on interpreting selected drivers.

\subsection{Influence of Sunspot Number - Timescale Matters} 
\label{sect:sunspotnumberaveragingeffect}

The measured or observed quantity for solar activity is the daily sunspot number. These observations display large variability as seen in figures~\ref{fig:fulltsOMNI} and ~\ref{fig:fulltsSTBSTA}. As we will demonstrate here, the stochasticity has an impact on the causal association with forecast accuracy term, MAE. Figure~\ref{fig:omnissnmae} shows the corresponding distribution of causal strengths of the pairs of MAE with 27-day smoothed {\em right} and daily unsmoothed {\em Left} SN and the latitudinal offset ($\dthta$). The unsmoothed daily sunspot number (SN) has lower cs ($=MI/H$) than the latitudinal offset ($\dthta$). However, upon performing a rolling mean on the daily SN to yield 27-day smoothed averaged SN or the $\dssn$, the hierarchy reverses. As shown in figure~\ref{fig:omnissnmae}, we see that the total causal strength of SN, $cs(\dssn\rightarrow MAE_{27})$, goes from 0.03 to 0.14 upon averaging, compared to $cs(\dthta\rightarrow MAE_{27})$ with a mean value of 0.05. As expected, the stage of solar cycle and overall time variability of the Sun is better represented by the smoothed SN, which has a significant influence on MAE. That the daily SN has a significant stochastic component is also confirmed by / evident from the entropy estimates. The entropy is reduced upon smoothing or averaging SN and is lower than that of $\dthta$ by a factor of a few (for example $\approx$ 2 for STB-STA ) ; however, this is not a significant effect.

\subsection{Conditional and Interaction Information : Higher order terms}
\label{sect:cmiii}

\begin{figure}    
     \centerline{\includegraphics[width=0.85\textwidth,clip=]{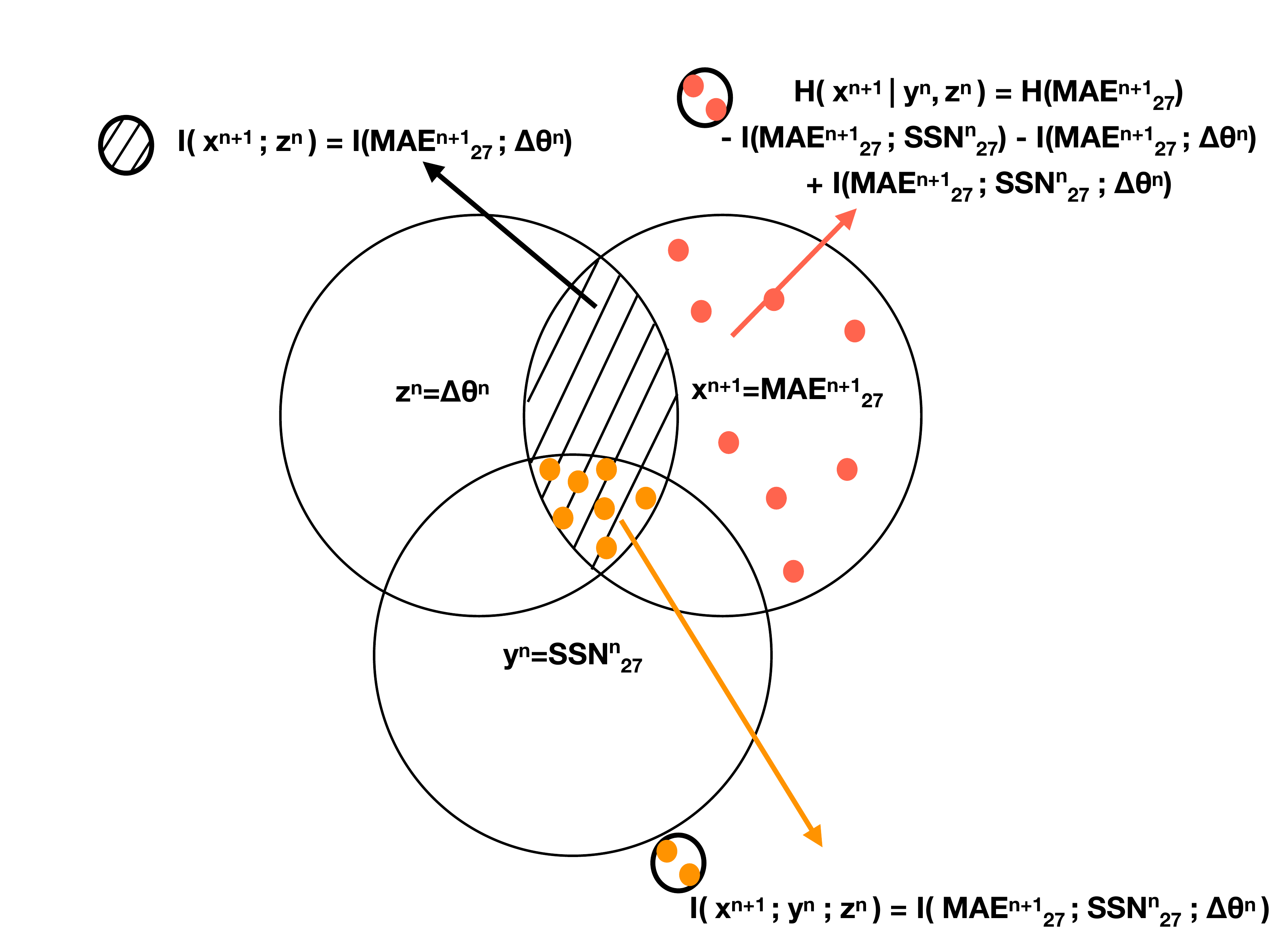}
              }
              \caption{The figure shows the different information components for three variables. The intersections represent information shared between variables. The black striped region is the mutual information shared between target MAE and driver $\dthta$. The yellow dots denote interaction information : information shared between $MAE_{27}$ and both $\dthta$ and $\dssn$. The pink circles show the entropy or uncertainty in $MAE_{27}$ that is not explained or shared by either $\dssn$ or $\dthta$ or their interaction.}
   \label{fig:intvenn}
   \end{figure}

   In presence of multiple causes or drivers (say y and z), the aforementioned causal strength term, $cs(x ; z)$ will come to represent the fractional reduction in uncertainty in the target due to the driver, z. And there's a similar term for y. To  further disentangle and isolate the influence of each driver we also compute the conditional mutual information (CMI), e.g., $I(x ; z | y)$. For two drivers y and z (and a single target x), conditional mutual information, $I(x ; z | y)$ given in eqn~\ref{eqn:condinfdefn}, `conditions out' the effect of one driver ($y$), leaving the direct influence of the other one ($z$). This can be visualised in terms of Venn diagrams in figure~\ref{fig:intvenn}. It is the difference between the intersection of x and z circles (black stripes) and that of x, z and y circles (yellow spots). In our application to corotation forecasts, the example used for illustration has x as $MAE_{27}$ and the y and z as the drivers $\dthta$ and $\dssn$, respectively. We will keep the same normalisation with entropy for all information terms so that they can be combined or compared. 

\begin{figure}    
\centerline{\includegraphics[width=0.85\textwidth,clip=]{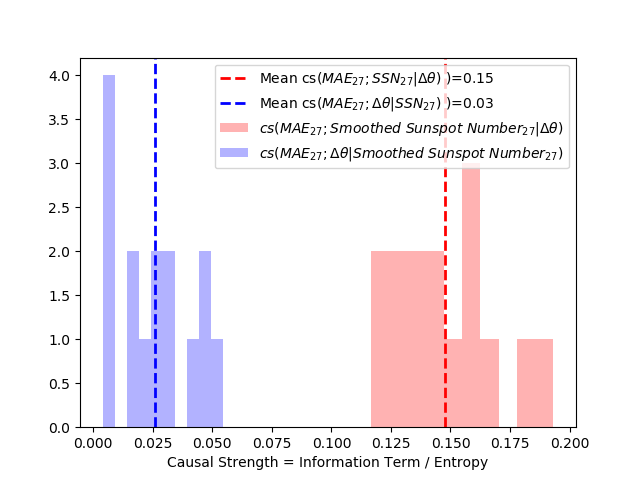}
              }
              \caption{The figure shows the {\em Left:} distribution of conditional causal strengths with OMNI data. cs here is associated with conditional mutual information normalised by the entropy, H, for the combinations of $MAE_{27}$ with Sunspot Number ($\dssn$) (27 day rolling averages for both) and Latitudinal Offset for the full dataset, namely $cs(MAE_{27} ; \dssn | \Delta \Theta)$ and $cs(MAE_{27} ; \Delta \Theta | \dssn)$.}
   \label{fig:latoffsetssncmi}
   \end{figure}

\begin{figure}    
     \centerline{\includegraphics[width=0.85\textwidth,clip=]{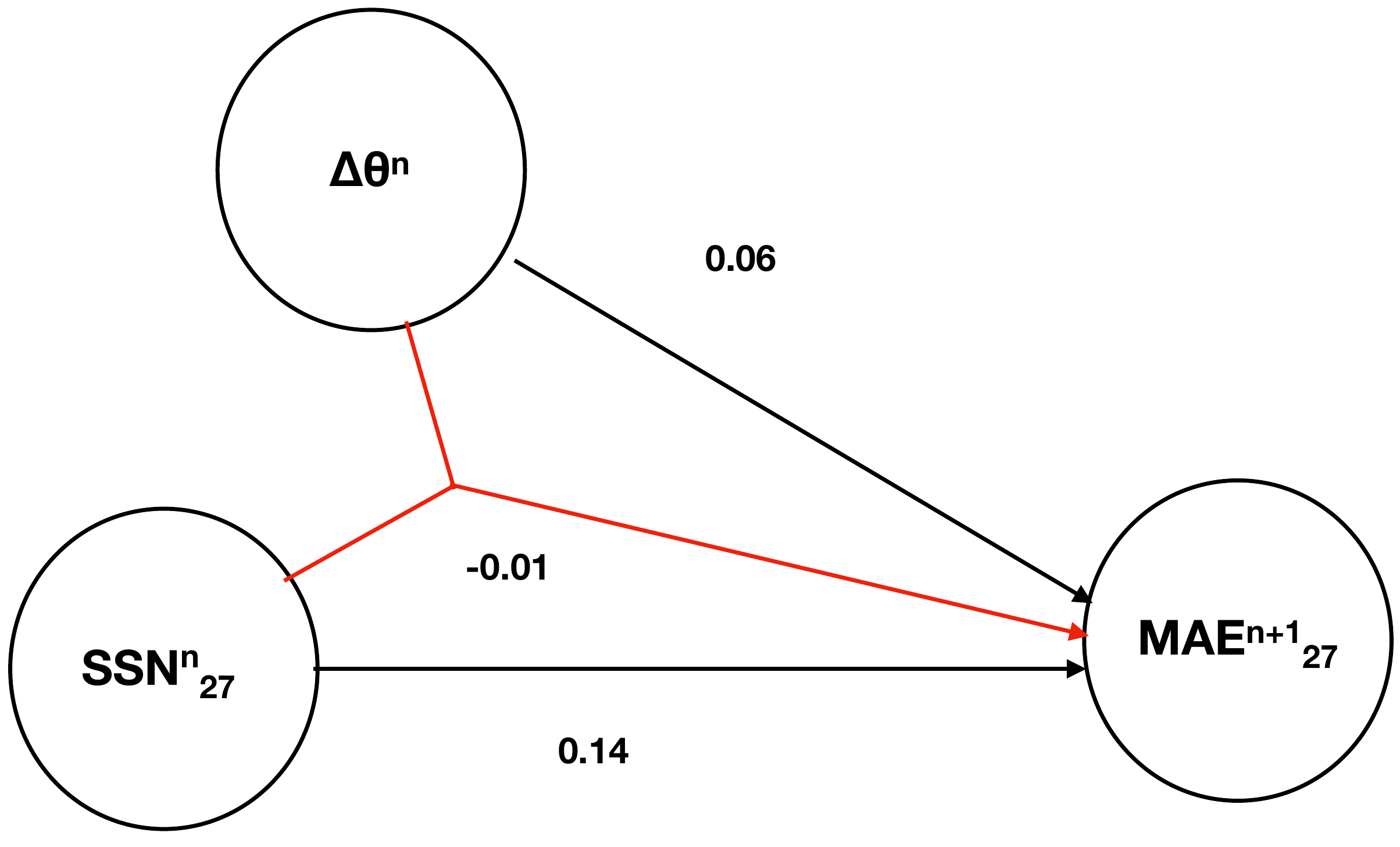}
              }
              \caption{The figure summarises the mean $MAE_{27}$ dependence on  latitudinal offset $\dthta$ and $\dssn$ individually (black) and in combination (red) for OMNI-OMNI. This is done in terms of the causal strength ($cs=\frac{\rm Information\ Term}{\rm Entropy}$) values defined earlier - the numbers attached to the arrows. The superscripts merely indicate the formal need for the causes ($\dssn^{n},\dthta^{n}$) to precede the effect ($MAE_{27}^{n+1}$) - in practice for this application, a single time-step makes negligible difference.}
   \label{fig:causalsummdiagOMNIOMNI}
   \end{figure}

The conditional mutual information can be defined in terms of the conditional entropies as,
\begin{equation}
\label{eqn:condinfdefn}
    I(x ; y | z) = H(x | z) - H(x | y, z)
\end{equation}
The above equation for the conditional mutual information of x with respect to y and z represents the difference in entropy of x ``conditioning out'' z alone ($H(x | z)$) and entropy of x ``conditioning out'' y and z together ($H(x | y, z)$). This leaves us with the direct influence of driver y on target x, excluding any indirect influence mediated by or shared with z.  The distributions of such conditional information terms for the triplet ($MAE_{27}$,$\dthta$,$\dssn$) are estimated in figure~\ref{fig:latoffsetssncmi} ; these provide the so-called direct causal influence contribution of $\dssn$ and $\dthta$ on $MAE_{27}$. These are symbolised by the black arrows in the causal summary diagram in figure~\ref{fig:causalsummdiagOMNIOMNI}. The causal summary diagram, as the name suggests, provides a summary of the information flow from (and therefore the causal influence of) the driver variables ; in this case the latitudinal offset ($\dthta$) and the smoothed sunspot number ($\dssn$).  Now the interaction information can be written in terms of the mutual and conditional mutual information as,
\begin{eqnarray}
\label{eqn:intinfdefn}
    I(x ; y ; z) &=& I(x ; y) - I(x ; y | z) \nonumber \\
    &=& I(x ; z) - I(x ; z | y)
\end{eqnarray}
This equation for the interaction information of x with y and z gives the difference between the mutual information shared between x and y ($I(x ; y)$) and information shared between them, upon conditioning out z ($I(x ; y | z)$). This is the interaction information shared between the 3 variables, x, y and z and is symmetric in all three variables. If we fix one as the target with the other two as drivers, as we do for our application, then the expression for interaction information is symmetric in the two drivers as demonstrated by the two equivalent expressions for $I(x ; y ; z)$ in equation~\ref{eqn:intinfdefn}. So it doesn't matter which driver we condition on. We will exploit this later on as estimates from actual measurements may not converge to the same value as eqn~\ref{eqn:intinfdefn}. So we can take the average of the two symmetric expressions to represent the interaction information between one target and two drivers. This is seen later in the observational estimates. 

This quantity can be interpreted as the information shared between x and y, less the information shared between them when z is known. If the interaction information is non-negative,  or $I(x ; y) \geq I(x ; y | z)$, it implies that the dependency of x on z partially or entirely (equality) constitutes the dependency on y \citep{GhassamiK17}. If the interaction information is negative, or  $I(x ; y) < I(x ; y | z)$, then each one of the variables induces and increases correlation between the other two. 

In the previous subsection, we ascertained that the smoothed sunspot number or $\dssn$ is more appropriate as a proxy for the solar activity in evaluating its causal influence on the average corotation forecast accuracy, $MAE_{27}$. Now we wish to disentangle the direct and indirect effects of both $\dssn$ and the latitudinal offset, $\dthta$ on $MAE_{27}$. Their joint effect, or one mediating through the other, is naturally a higher order effect and hence we need the higher order information terms. We compute the higher order information theoretic quantities, namely conditional mutual information and interaction information between drivers $\dssn$ and $\dthta$ and the target, $MAE_{27}$, still using only the OMNI dataset. 
For $MAE_{27}$ (x), $\dthta$ (y) and $\Delta t$ (z), the interaction information corresponds to the common part with yellow circles in the Venn diagram in figure~\ref{fig:intvenn}. This is therefore the information shared across all three variables in general.

Formally, causality necessitates there be a time lag between the cause and effect such that the former precedes the latter. And indeed there is a time lag between the forecast accuracy of a future step, $MAE_{27}^{n+1}$, and the drivers $\dthta^{n}$ and $\dssn^{n}$. This is innate/intrinsic to the way the time-series observations are done. However, in this particular application, we are considering daily variations and the drivers -- latitudinal separation, longitudinal separation and 27-day smoothed sunspot number -- vary over much longer timescales. Thus a single time-step between n and n+1 makes negligible difference to the computed information components. However, the notation involving target at n+1 and drivers at n is maintained to demonstrate the general principle.

\subsection{Symbolic representation : Causal Summary diagrams}
The causal information flow between the variables is summarised in figure~\ref{fig:causalsummdiagOMNIOMNI}. The nodes (or ovals) represent the variables and the arrows represent the flow of information to the target variable, $MAE_{27}$ one time step in the future (n+1). Black arrows represent the influence of single driver conditioning out influence of the other drivers. The red segments ending in an arrowhead on the target represents the joint causal influence of the drivers. The confluence of the segments out of the drivers into a point symbolises this join or combined effect ; the arrowhead as usual points to the information flow into the target. This represents the component of influence that is driven by the combination of drivers together, distinct from their individual, direct influences on the target, shown by black arrows. This combined or joint effect could be a positive one showing a redundancy in driver or that one driver partially or entirely captures the influence due to another. It could be negative suggesting that one driver induces an influence from the other driver. These can be mathematically quantified in terms of the interaction information. Here for 27-day corotation forecasts using only OMNI data, the only drivers are $\dthta$ and $\dssn$, now considered simultaneously. (As we will see in an upcoming section, the lead time $\dlt$ -- related to the longitudinal separation -- will have a role to play for STEREO data.) We find that the direct influences of $\dssn$ and $\dthta$ (black arrows) are more important than the joint influence (red arrow) on $MAE_{27}$. In general, we can compute joint influence due to multiple drivers starting from pairs (the red arrows) to the joint influence of all $n$ drivers simultaneously. However, to use full general mathematical framework in \cite{vanleeuwen2021framework} is computationally expensive and complex. It is also not essential in our work here to get the main dependencies. We compute the causal strengths (defined earlier) from the mutual and conditional mutual information terms in accordance with \citet{vanleeuwen2021framework}. The black arrows are given by: 
\begin{equation}
 \large( \dthta^{n} \rightarrow MAE_{27}^{n+1} \large)_{1link} = I(MAE_{27}^{n+1} ; \dthta^{n} | \dssn^{n})   
\end{equation}

\begin{equation}
\large( \dssn^{n} \rightarrow MAE_{27}^{n+1} \large)_{1link} = I(MAE_{27}^{n+1} ; \dssn^{n} | \dthta^{n})
\end{equation}
The red arrow symbolising the joint influence of $\dthta$ and $\dlt$ represents and is related to the interaction information shown in the figure~\ref{fig:intvenn}. Graphically this represents the intersection of the information component common to each of the three variables in our triplet i.e. two drivers (latitudinal offset and lead time) and target ($MAE_{27}$ ). This is therefore symmetric and is, theoretically, independent of the variable that it is conditioned on. However, when estimating from measured quantities, this symmetry, indicated both graphically in figure~\ref{fig:intvenn} and equation~\ref{eqn:intinfdefn} is not strictly adhered to. Hence, we can express the joint influence indicated by the red arrow as the average of the two equivalent ways of estimating it as:

\begin{eqnarray}
\label{eqn:redarrowlatoffssn}
 &\large( \dthta^{n}& \rightarrow MAE_{27}^{n+1} \large)_{2link} + \large( \dssn^{n} \rightarrow MAE_{27}^{n+1} \large)_{2link} \nonumber \\
&=& 1/2\ \large[\ I(MAE_{27}^{n+1} ; \dthta^{n}) - I(MAE_{27}^{n+1} ; \dthta^{n} | \dssn^{n}) \nonumber \\
&+& I(MAE_{27}^{n+1} ; \dssn^{n}) - I(MAE_{27}^{n+1} ; \dssn^{n} | \dthta^{n}) \large]
\end{eqnarray}

     \begin{figure}    
   \includegraphics[width=0.8\textwidth,clip=]{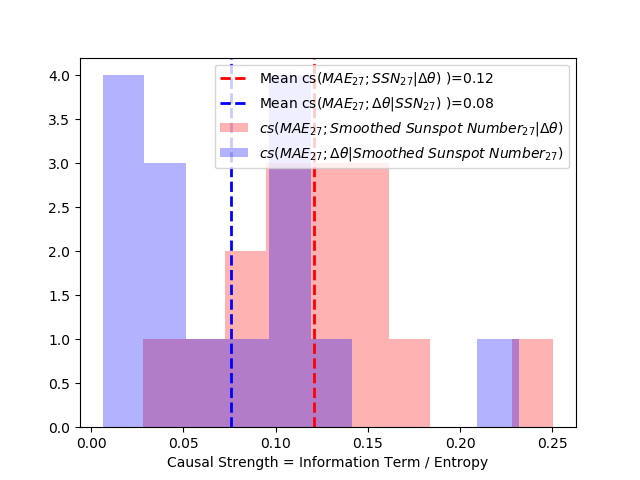}
   \includegraphics[width=0.8\textwidth,clip=]{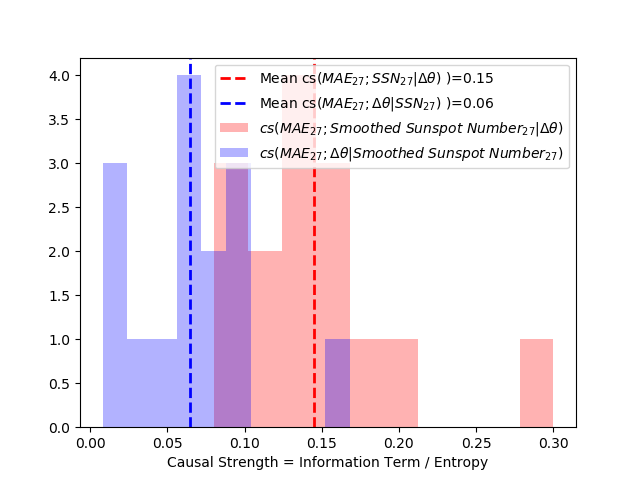}
   \includegraphics[width=0.8\textwidth,clip=]{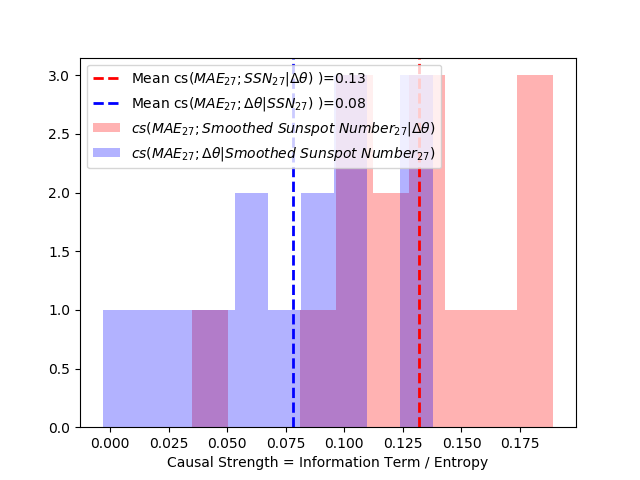}
              
              \caption{$MAE_{27}$ dependence on principle drivers for corotation forecasts;  latitudinal offset $\dthta$ conditioned on smoothed sunspot number or $\dssn$ and vice versa in pairs of {\em Top:} STB-STA, {\em Middle:} STB-OMNI and {\em Bottom:} OMNI-STA. }
   \label{fig:STBSTAOMNIconsistency}
   \end{figure}

\subsection{Consistency across Datasets}
\label{omnistbstaconsistency}

We next test this relative influence of $\dssn$ and $\dthta$ on $MAE_{27}$ across the available datasets, namely STB-STA, STB-OMNI and OMNI-STA pairings. This is shown in figure~\ref{fig:STBSTAOMNIconsistency}. In each case, we find the interaction information, $I(MAE_{27}^{n+1} ; \dthta^{n} ; \dssn^{n}$) to be positive. This is an indication that $\dssn$ partially constitutes the dependency of $MAE_{27}$ on $\dthta$ and vice versa, but it is not very significant. And across these three datasets (as well as OMNI-OMNI recurrence forecasts), we found that the direct causal strengths of latitudinal offset, $I(MAE_{27}^{n+1} ; \dthta^{n} | \dssn^{n})$, is around $60-70 \% $ of the direct causal strength of $\dssn$,  $I(MAE_{27}^{n+1} ;  \dssn^{n} | \dthta^{n})$. Furthermore, estimates of the interaction information, given by $I(MAE_{27}^{n+1} ;  \dssn^{n}) - I(MAE_{27}^{n+1} ;  \dssn^{n} | \dthta^{n})$, are merely $\sim 7 \% $ of the direct causal influence, as was also shown in OMNI dataset in figure~\ref{fig:causalsummdiagOMNIOMNI}.  This suggests that to a good approximation, the causal influence of the solar activity is decoupled from that of the latitudinal offset. This will aide us in considering the causal influence of lead time in turns with these two variables, simplifying the causal network.

\section{STEREO : Effect of Lead Time}
\label{sect:stereohigherorder}
As explained in the previous section, the OMNI (27-day) corotation forecast dataset allows us to focus on the causal influence of $\dthta$ and $\dssn$ as proxy of the solar activity on $MAE_{27}$. Having learnt that the interaction information between these three driver variables is small, we can assume their influence to be largely independent. We will now proceed to pair $\dthta$ and $\dssn$ by turns, with the lead time $\dlt$. This will give us the direct and interaction terms for each case, analogous to the causal network in figure~\ref{fig:causalsummdiagOMNIOMNI}. 
     \begin{figure}    
     \centerline{
     \includegraphics[width=0.5\textwidth,clip=]{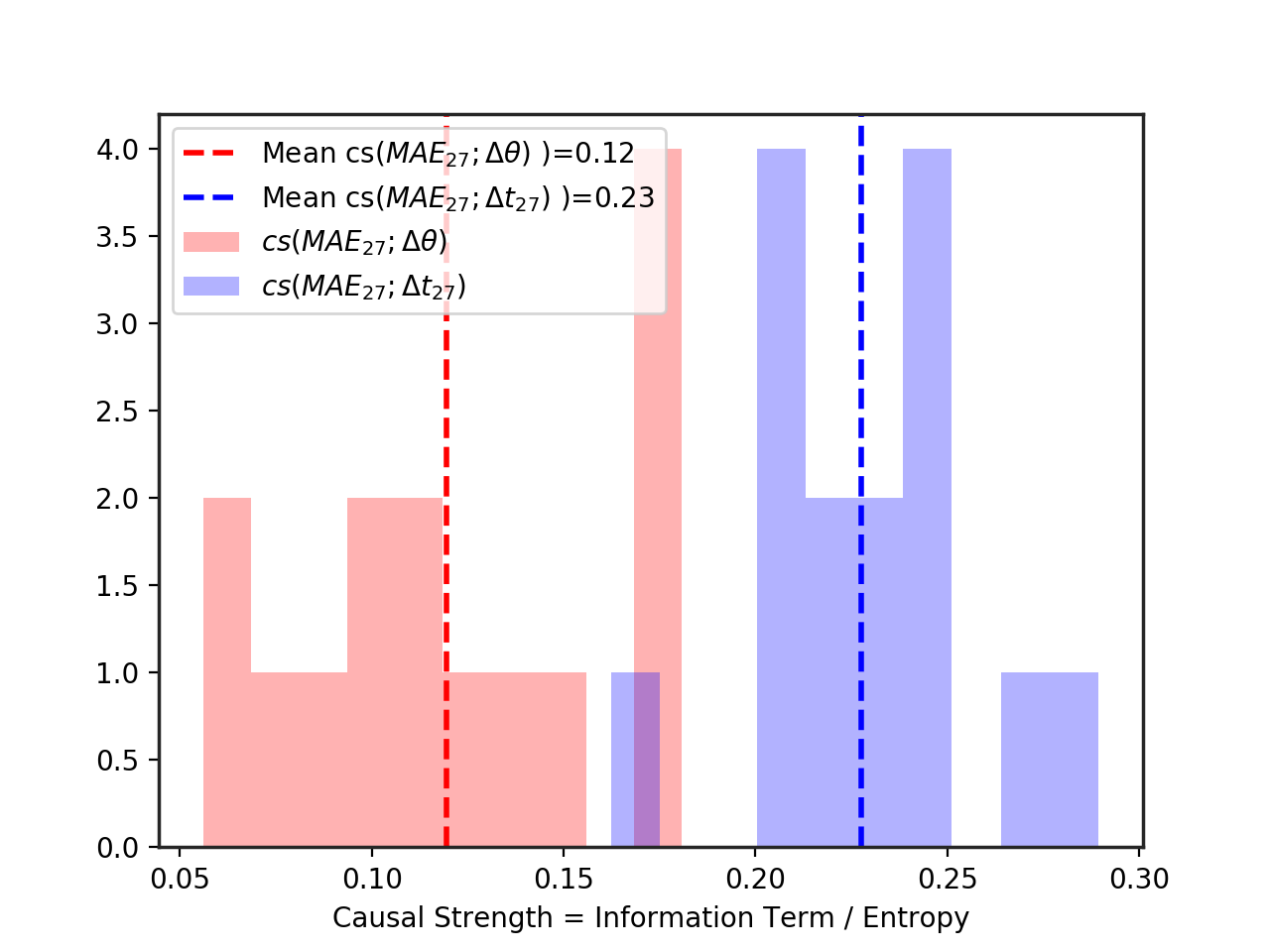}
     \includegraphics[width=0.5\textwidth,clip=]{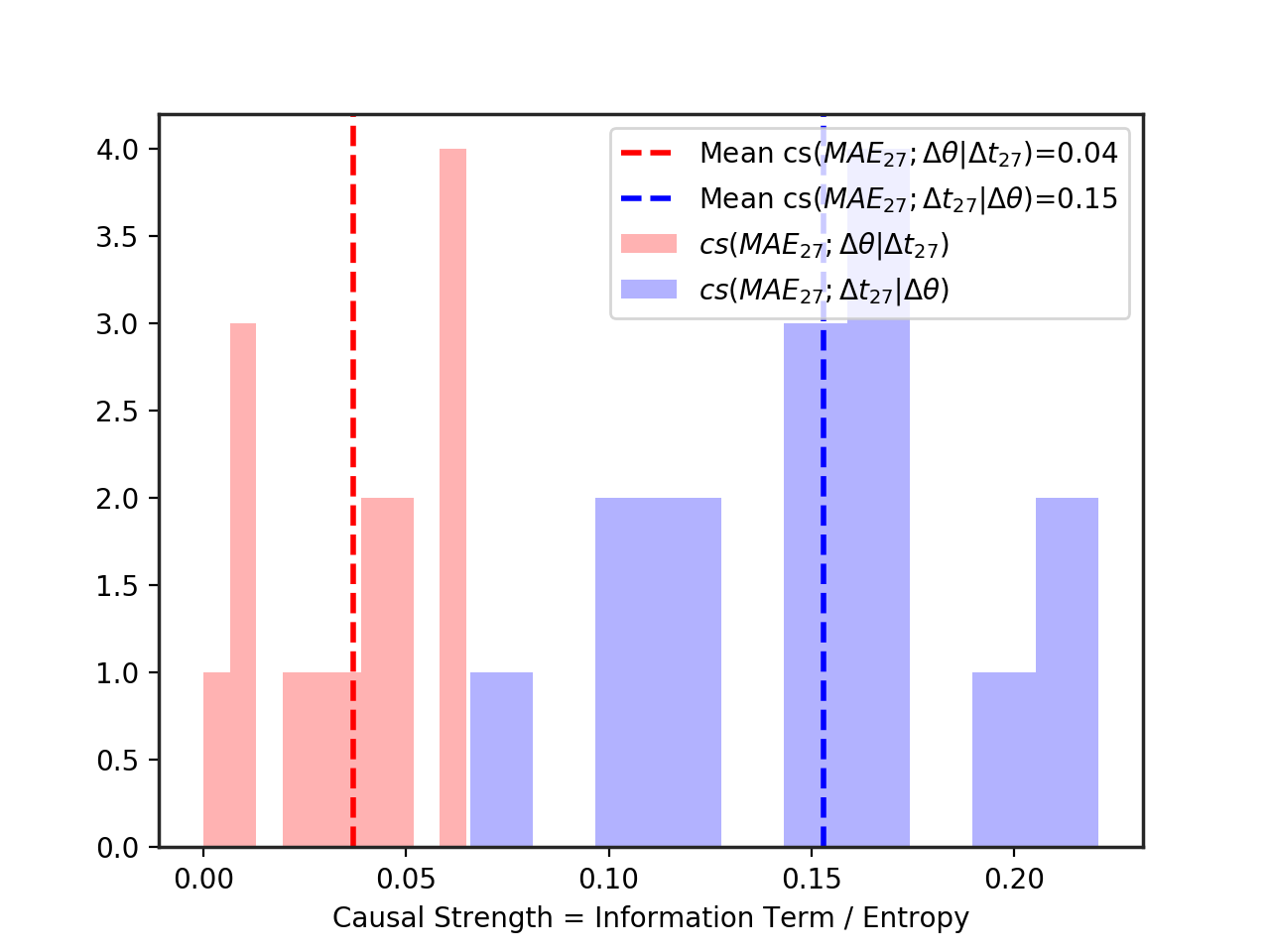}
     }
              \caption{The figure shows the histogram of causal strengths, cs, of drivers -- the lead time, $\dlt_{27}$ and $\dthta$ -- with the target, $MAE_{27}$ using STA-OMNI data. On the {\em Left:}, are the distributions of pairwise strengths (MI/H) and on the {\em Right:} are the conditional causal strengths (CMI/H).}
   \label{fig:staomnimaelatofflead}
   \end{figure}
\subsection{Conditional Causal Influence : Lead Time}

Analogous to the causal analysis of the OMNI time-series, we estimate causal linkages using the STA-OMNI corotation time-series. We begin with the time-series of $\dlt$ and $\dthta$ as drivers along with the $MAE_{27}$ as the target. We compute causal strength terms corresponding to mutual information terms $I(MAE_{27}^{n+1} ; \dthta^{n})$ and $I(MAE_{27}^{n+1} ; \dlt^{n})$ and the conditional information terms, $I(MAE_{27}^{n+1} ; \dthta^{n} | \dlt^{n})$ and $I(MAE_{27}^{n+1} ; \dlt^{n} | \dthta^{n})$. 

$I(MAE_{27}^{n+1} ; \dthta^{n} ; \dlt^{n}$) is actually the influence of $\dthta$ on $MAE_{27}$ that is shared by (or mediated through / induced by)  $\dlt$, in general. From figure~\ref{fig:staomnimaelatofflead}, it is evident from the histograms of $cs(MAE_{27}^{n+1} ; \dthta^{n})$ and $cs(MAE_{27}^{n+1} ; \dthta^{n} | \Delta t^{n})$ (or equivalently the corresponding information terms) that the interaction information, $I(MAE_{27}^{n+1} ; \dthta^{n} ; \Delta t^{n})$ is positive. Using the mean values of each term in the information triplet, direct (i.e. conditional mutual) and interaction information for ($MAE_{27}$,$\dthta$,$\dlt$) we get an interaction information amounts to $\sim 20\%$. The direct influence of $\dlt$ on $MAE_{27}$ is $\sim 50 \%$ and the direct influence of $\dthta$ is the remaining $\sim 30\%$. These numbers are the relative proportions of influence of the two drivers considered $\dthta$ and $\dlt$, without considering $\dssn$. The `missing' $20\%$ is likely to have a significant contribution from the other driver, $\dssn$. However, we also have noise, which includes the statistical fluctuations. Also, the averaged sunspot number is used as a proxy for the time evolving nature of the solar wind ; it's not an exact proxy.

\begin{figure}    
     \includegraphics[width=0.8\textwidth,clip=]{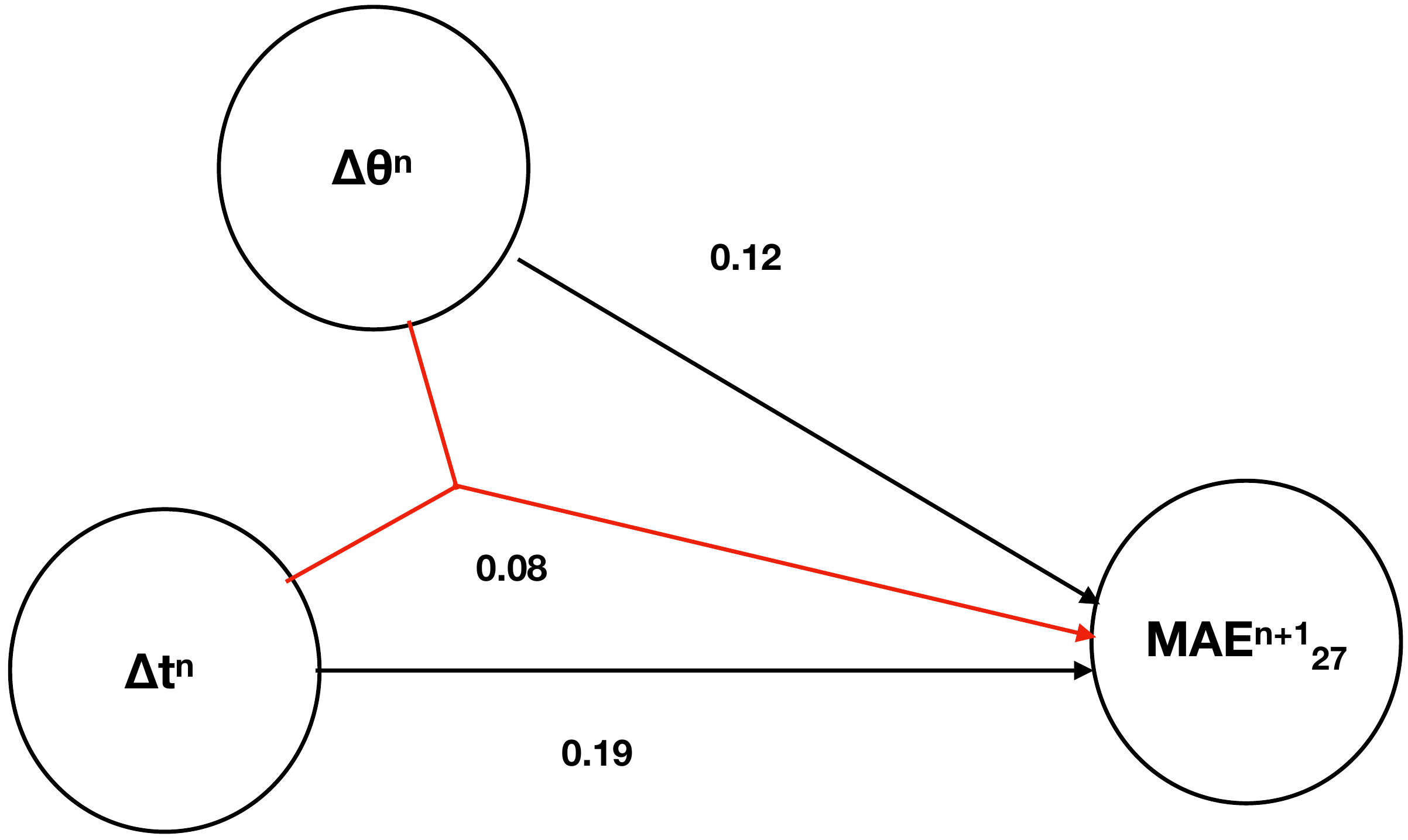}
    \includegraphics[width=0.8\textwidth,clip=]{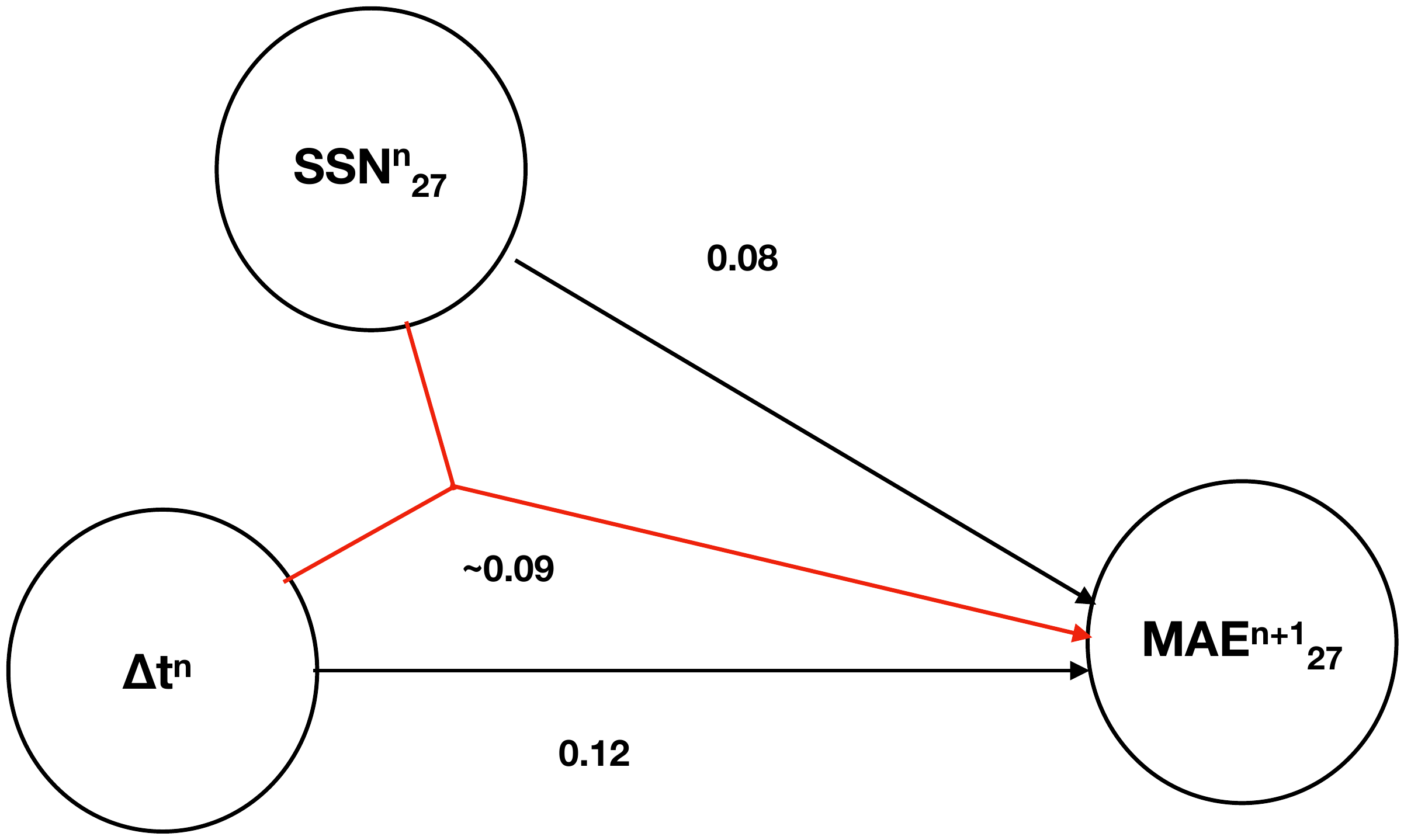}
     \includegraphics[width=0.8\textwidth,clip=]{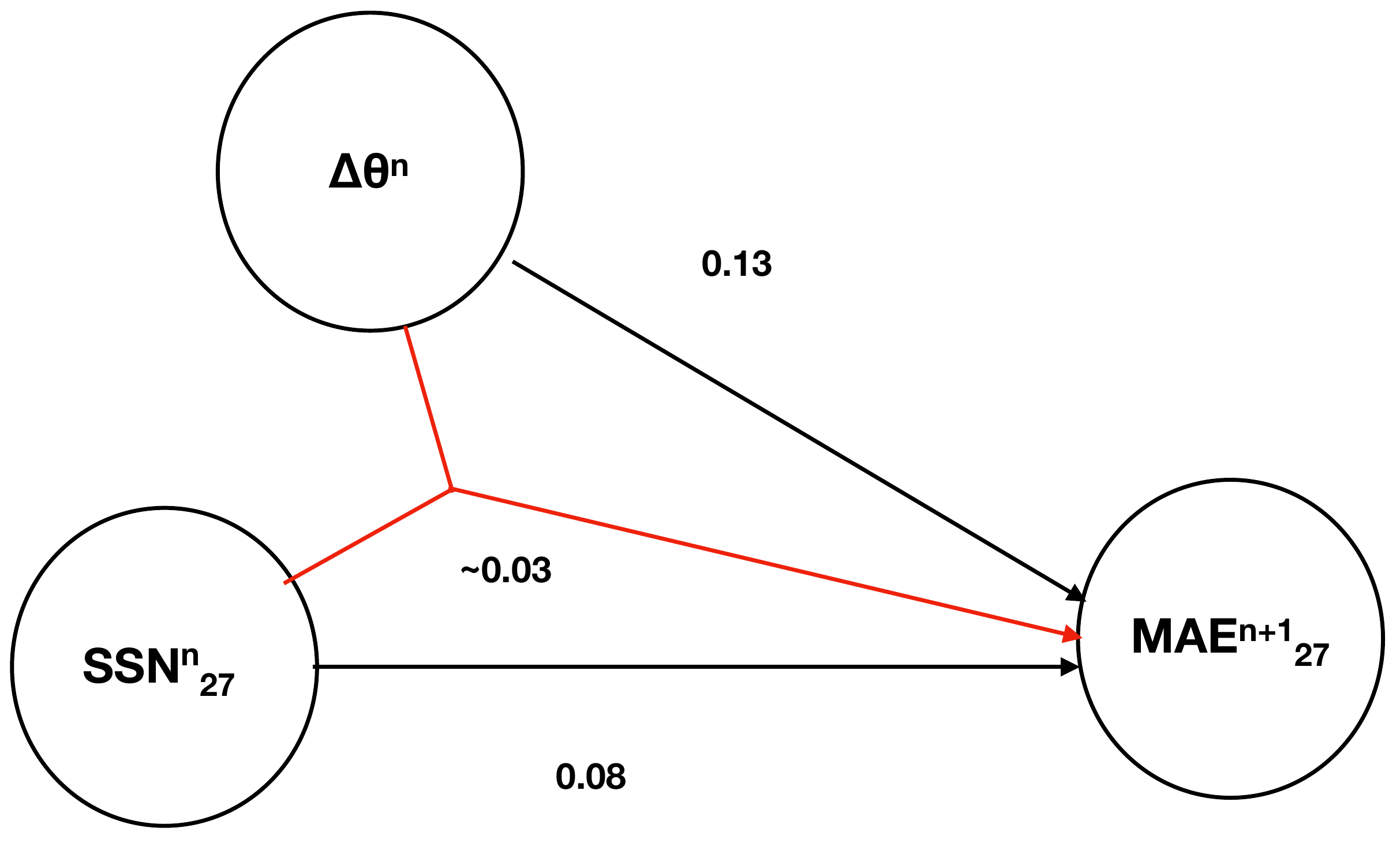}
              \caption{The figure summarises the mean causal dependence of forecast accuracy $MAE_{27}$ on latitudinal offset $\dthta$, lead time $\dlt$ and average/smoothed sunspot number $\dssn$ individually (black) and in pairwise combinations (red) for STA-OMNI.}
   \label{fig:causalsummdiagSTAOMNI}
   \end{figure}

Histograms with 15 bins as before are used to estimate the distribution of these causal strength terms and shown in figure~\ref{fig:staomnimaelatofflead}. It is clear that the lead time, $\dlt$, has an influence on the $MAE_{27}$. Upon conditioning on $\dthta$, we find a sizable direct influence of $\dlt$ on $MAE_{27}$, around 1.5 times greater than the direct influence of $\dthta$ ; this is shown by the black arrows in the summary diagram in figure~\ref{fig:causalsummdiagSTAOMNI}.

Next we look at the driver pair of lead time and (smoothed or rolling 27 day average) sunspot number $(\Delta t,\dssn)$. We again estimate the pairwise (mutual information) and direct dependencies (conditional mutual information) of $\dlt$ and $\dssn$ on $MAE_{27}$. Once again, the two drivers $\dssn$ and $\dlt$ have shared dependencies. This is summarised in figure~\ref{fig:causalsummdiagSTAOMNI}. 
   
   The black arrows are given by : 
\begin{equation}
 \large( \dlt^{n} \rightarrow MAE_{27}^{n+1} \large)_{1link} = I(MAE_{27}^{n+1} ; \dlt^{n} | \dssn^{n})   
\end{equation}
and
\begin{equation}
\label{eqn:blackarrowleadtimessn}
\large( \dssn^{n} \rightarrow MAE^{n+1} \large)_{1link} = I(MAE^{n+1} ; \dssn^{n} | \dlt^{n})
\end{equation}
The red arrows symbolising the joint influence can be written symmetrically as a sum of 
\begin{eqnarray}
\label{eqn:redarrowleadtimessn}
 &\large( \dlt^{n}& \rightarrow MAE_{27}^{n+1} \large)_{2link} + \large( \dssn^{n} \rightarrow MAE_{27}^{n+1} \large)_{2link} \nonumber \\
&=& 1/2\ \large[A\ I(MAE_{27}^{n+1} ; \dlt^{n}) - I(MAE_{27}^{n+1} ; \dlt^{n} | \dssn^{n}) \nonumber \\
&+& I(MAE_{27}^{n+1} ; \dssn^{n}) - I(MAE_{27}^{n+1} ; \dssn^{n} | \dlt^{n}) \large]
\end{eqnarray}

The quantitative analyses is summarised in the causal summary diagram in fig~\ref{fig:causalsummdiagSTAOMNI}. The direct causal influence of $\dlt$ and $\dssn$ are $\approx 41 \%$ and $\approx 28 \%$, respectively, in relative terms. And the joint influence is $\approx 31\%$.

 \subsection{Dependencies of the driver triplet}  
   \label{sect:drivertriplet}
  \begin{figure}    
     \includegraphics[width=0.70\textwidth,clip=]{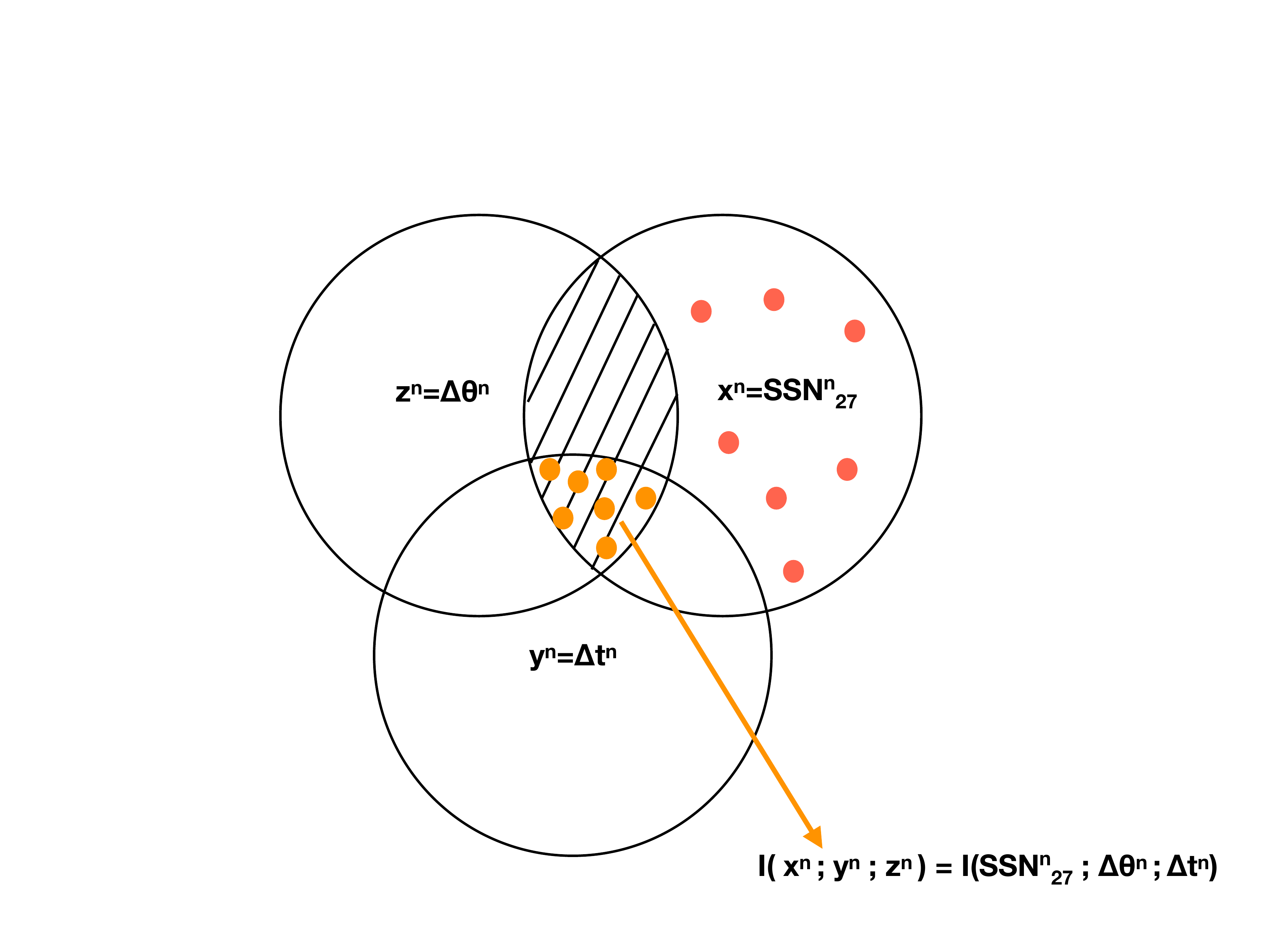}
     \includegraphics[width=0.70\textwidth,clip=]{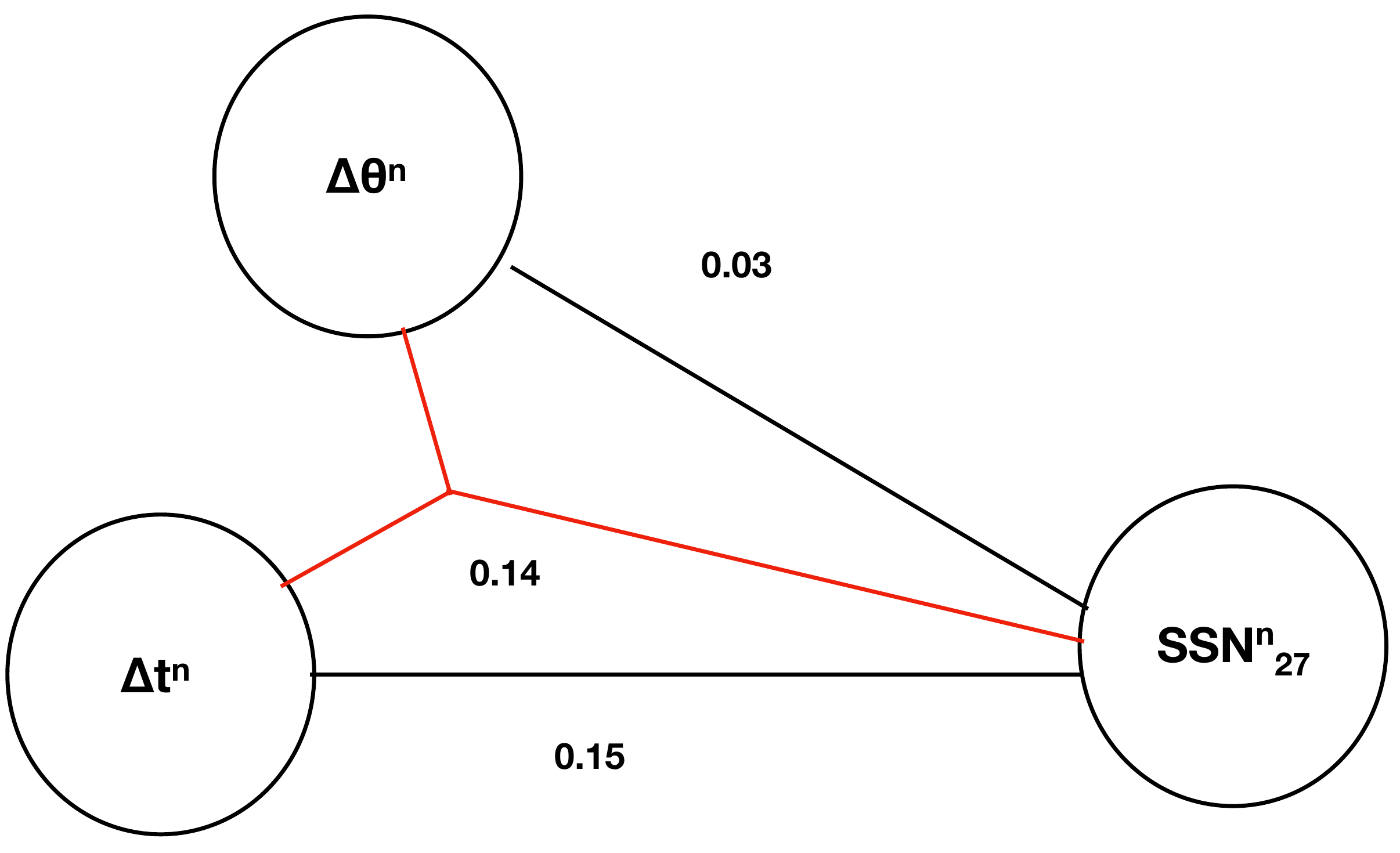}
              
              \caption{The figure summarises the mean (in)dependence of the smoothed sunspot number ($\dssn$) on the latitudinal offset $\dthta$ and corotation time $\dlt$ individually (black) and in combination (red) for STB-OMNI. Here we test the interdependence of the drivers and not a relation between causes preceding the effect symbolically shown 
 by the absence of the arrow heads ; hence each is at n.}
   \label{fig:causalsummdiagSTBOMNIdrivers}
\end{figure}
  Here we put aside the target variable, $MAE_{27}$, and apply the causal measures to explore the dependencies between the drivers themselves. The goal is to directly probe the statistical (in)dependence of the drivers without any lag i.e. we do not seek causal information flow in time, from one variable to another. Instead we look simply for information overlap which tells us the how the drivers relate to each other. Hence, the three variables are $\dssn^{n}$, $\dthta^{n}$ and $\dlt^{n}$. These dependencies between drivers are symbolised by line segments instead of arrows. So with no natural target variable, we treat $\dssn$ effectively as the target. The reason for this, is that apriori we might expect a stronger dependence between $\dthta^{n}$ and $\dlt^{n}$ and hence we wish to see how these two affect $\dssn$. Our approach is reflected in the causal diagram shown in figure~\ref{fig:causalsummdiagSTBOMNIdrivers}. We find that the direct effect (black line segment) of $\dlt$ on $\dssn$ is greater than that of $\dthta$ by over an order of magnitude. This is because the solar activity does depend upon the phase and hence the lead time across cycles. And while the joint effect (red segment) of $\dlt$ and $\dthta$ on $\dssn$ is lower than the direct effect of $\dlt$, it is still non-trivial. As the corresponding interaction information term is positive, it suggests that $\dlt$ mediates the dependency on $\dthta$.

\section{Conclusion} 
      \label{S-Conclusion} 

In this paper, we probe what drives the  accuracy of corotation forecasts of the solar wind observations. As we do not have means to make interventional experiments, we apply causal inference methods to the available observations. The causal drivers of relevance are the latitudinal separation or offset ($\dthta$) between the observing and forecast locations, the forecast lead time ($\dlt$) and the solar activity, as measured by 27 day rolling average of the sunspot number ($\dssn$). The target variable is the forecast accuracy measured in terms of the mean absolute error between observation and forecast. We use information theoretic measures to estimate the strength of the causal influence of the drivers on the target. These do not merely estimate correlations between pairs of variables, but can disentangle the influence of a third variable via conditioning the information content shared between three of them. Depending on the different information terms or components, we can estimate the influence of the individual drivers as well as their joint effect, induced effects, and redundancy due to correlation amongst drivers. 

We draw the following conclusions:
\begin{enumerate}
    \item The decomposition of information flow between the different drivers (or causes) and the target is effective in identifying cause and effect relationships driving dynamical systems in presence of complex, non-linear relationships between multiple variables. This approach is perfectly suited to trace what drives the corotation forecast accuracy or rather the uncertainty. 
    \item The pairwise causal relationship between drivers -- latitudinal offset ($\dthta$), forecast lead time ($\dlt$), and the average sunspot number ($\dssn$) -- and target $MAE_{27}$ given by mutual information normalised to the target entropy confirms our understanding from previous results \citep[for eg.][]{turner2021corot}. The solar activity levels measured via $\dssn$ and the lead time $\dlt$ have a big influence on the average forecast error, $MAE_{27}$, followed by the latitudinal offset, $\dthta$. Statistical noise levels impacts the absolute values of the dependencies. The relative values of the causal strengths, however, teach us of the hierarchy of influence of the different driver combinations.
    \item Exploiting higher order measures like interaction information, we can probe deeper into  causal influence of multiple drivers in conjunction with one another. A non-zero interaction information has two possible cases with corresponding interpretations. Negative values show an induced effect (i.e. one driver showing a coupling to the target induced only due to the presence and influence of another driver) whereas positive values show a redundant / shared influence. We can be quantitative about the relative importance of joint causal (positive or negative) influence and therefore potentially impact forecasting. Also qualitatively, knowledge of whether drivers act independently of one another, can help design future data experiments. For instance, from fig.~\ref{fig:STBSTAOMNIconsistency}, it is clear from the consistency of $MAE_{27}$ dependence on $\dssn$ and Latitudinal Offset -- from OMNI alone and OMNI with STEREO -- that solar activity has a stronger, direct influence on $MAE_{27}$ than latitudinal offset, independent of lead time. Hence, the appropriate weight can be given to each of these drivers in an assimilation forecast. On the other hand, from  fig.~(\ref{fig:causalsummdiagSTBOMNIdrivers}) while solar activity and latitudinal offset have a weaker direct association (black line between them) relative to its stronger, direct coupling with lead time,they do have an indirect coupling. The association of sunspot number with latitudinal offset is owed predominantly to the lead time. This implies, that the phase of the solar cycle is important. 
    \item The interaction information terms, such as $I(MAE_{27}^{n+1} ; \dthta^{n} ; \Delta t^{n})$ and \\ $I(MAE_{27}^{n+1} ; \dssn^{n} ; \dthta^{n})$ or equivalently the corresponding causal strength terms (I/H) quantify the information content shared between the 3 variables. From these terms, we can learn that the $\dssn$ and $\dthta$ share very little information. i.e. their influence on $MAE_{27}$ is predominantly independent of one another. On the other hand, for the STEREO dataset, $\dthta$ and $\dlt$ share a non-trivial fraction ($\sim 20\%$) of the their total information content (or influence on $MAE_{27}$). And in this case, the +ve sign of the interaction information indicates that $\dthta$ partially contributes to the influence of $\dlt$ on $MAE_{27}$, and vice versa. These effects could not be revealed by standard correlation analyses.
\end{enumerate}
      
Using a causal inference approach, disentangles the drivers of the forecast accuracy in ways that standard statistical analyses or data assimilation cannot. Rather one can improve upon these latter two, using causal diagnostics. It allows us to not only disentangle individual sources of uncertainty in the forecast, but also calculate partial and complete redundancies in drivers of this uncertainty. These learning can potentially be applied to improve the solar wind data assimilation forecasts. 

\begin{acks}
This work was part-funded by the Science and Technology Facilities Council (STFC) grant numbers ST/R000921/1 and ST/V000497/1.
\end{acks}
\appendix

\bibliographystyle{spr-mp-sola}
\bibliography{sola_bibliography_example}  

\end{article} 

\end{document}